\documentclass[acmsmall]{acmart}

\usepackage{colortbl}
\usepackage{xcolor}
\usepackage[dvipsnames]{xcolor}
\usepackage{multirow}

\AtBeginDocument{%
  }

\acmISBN{978-1-4503-XXXX-X/18/06}

\begin{document}

\title[The Activist’s Guide to the Decentralized Social Universe]{The Activist’s Guide to the Decentralized Social Universe: A Framework for Exploring How Decentralized Social Networks Can Support Collective Action}

\author{Sybille L\'egitime}
\email{sybille_legitime@brown.edu}
\orcid{0009-0005-5882-6918}
\affiliation{%
  \institution{Brown University}
  \city{Providence}
  \state{Rhode Island}
  \country{USA}
}

\author{Harini Suresh}
\affiliation{%
  \institution{Brown University}
  \city{Providence}
  \state{Rhode Island}
  \country{USA}}
\email{harini_suresh@brown.edu}

\renewcommand{\shortauthors}{L\'egitime et al.}

\begin{abstract}
The overreaches of mainstream social media platforms have been extensively reported and studied. For activist communities, these platforms pose risks of surveillance, censorship, or erasure. Decentralized social networks (DSNs) serve as alternative online spaces that appear to prioritize values such as user privacy, free speech, and community control. However, the decentralized ecosystem is vast and complex, making it difficult for communities to understand how to best use these platforms for their organizing aims. We address this gap by proposing a conceptual framework for navigating the DSN landscape that defines core activist community needs---minimal overhead, community building and reach, on- and offline safety, and operational sustainability---and links them to concrete platform affordances such as resource efficiency, interoperability, and data ownership. We apply the framework to (1) evaluate and compare the sociotechnical tradeoffs of two contemporary DSNs (Mastodon and Bluesky), (2) understand broader community configurations that emerge across different DSN infrastructures and their implications for collective action, and (3) explore how two distinct activist communities facing infrastructural and political constraints might use the framework to find platforms that align with their needs. We conclude by reflecting on the theoretical promises of DSNs and the structural conditions that shape and constrain participation across them.
 
\end{abstract}

\ccsdesc[500]{Human-centered computing~Collaborative and social computing theory, concepts and paradigms}

\keywords{Decentralized platforms, ecosystem, conceptual framework, social computing, activist communities}

\maketitle

\section{Introduction}

Social media platforms have experienced several mass exoduses in recent years. Following Elon Musk's acquisition of Twitter---later rebranded X --- in October 2022, millions of users migrated to Meta's Threads, a mainstream competitor, and to Mastodon, its decentralized counterpart. After the United States presidential election in November 2024, hundreds of thousands of users left X within a few days to join Bluesky, a microblogging application with decentralized tendencies. Bluesky, spearheaded by former Twitter leadership, had also been gaining popularity in countries like Brazil, where users joined in droves following the nationwide ban of X. Finally, Meta's termination of its fact-checking program in January 2025, along with its announcement of reducing content moderation around political discourse, triggered yet another wave of user migration away from a mainstream social media platform.

Such a significant divestment reveals acute public awareness of corporate social media's encroachments on user freedoms and civil liberties. Although centralized social media platforms present themselves as neutral intermediaries whose main function is to facilitate online social relationships with minimal interference \cite{gillespie_politics_2010}, research on the centralized social media landscape underscore how corporations leverage their extensive reach to collect unprecedented amounts of personal data and to unilaterally decide on the usage of those data for algorithmically-mediated behavior modification \cite{reviglio_thinking_2020} and content sorting \cite{cotter_shadowbanning_2023, stewart_perfect_2022}.

This centralized governance paradigm has been shown to deepen power asymmetries and disproportionately impact marginalized groups \cite{birhane_algorithmic_2021, marjanovic_theorising_2022, mohamed_decolonial_2020}, who are constantly subjected to disinformation campaigns, censorship, erasure, and surveillance~\cite{schyff_duplicitous_2020, choksi_under_2024, olson_along_2023}. In particular, activist communities---our focus in this work---often rely on social media platforms for internal organizing and external communication \cite{lutherSocialMediaActivists2025}, yet find their ability to organize, mobilize, and safely communicate severely challenged \cite{klonick_new_2018, york_silicon_2021, lupienIndigenousMovementsCollective2020}.
Additionally, social media algorithms often exacerbate epistemic injustice by obscuring how content is surfaced or suppressed and 
delegitimizing knowledge claims advanced by platform users \cite{cotter_shadowbanning_2023, stewart_perfect_2022}. 
As a result, many researchers and communities have advanced strategies to pressure centralized platforms and regain some agency within the social medium by performing user audits \cite{shen_everyday_2021}, using data strikes and conscious data contributions \cite{vincent_data_2021,vincent_cscw}, and engaging in ``algorithmic gossip'' \cite{cotter_shadowbanning_2023, mniestri_algorithmic_2023, karizatAlgorithmicFolkTheories2021}. 

Beyond centralized platforms themselves, there exists an alternative online ecosystem of \textit{Decentralized Social Networks} (DSNs)\footnote{Also commonly referred to as ``Decentralized Online Social Networks '' (DOSNs).} that self-identify as bastions of resistance to centralized online rule, and appear to share values of security, privacy, censorship resistance, and user independence. The DSN landscape encompasses diverse platforms, protocols, and communities, each with distinct properties and idiosyncrasies. 
Despite its potential, the DSN landscape remains difficult to navigate for those outside of academic and developer circles. While activist communities might want to divest from the overreach of centralized platforms (or face infrastructural or sociopolitical constraints that require them to do so), this decentralized ecosystem is technically complex and inconsistently documented.  
For instance, as Meta slashes its hate speech policies and DEI team \cite{tenbarge_metas_2025}, how might activist communities understand which (of the many available) alternative platforms offer spaces for self-governance, mutual help, and reduced corporate control that best suit their needs? The space of DSNs is vast and complex, as are the desires and concerns of different communities.

Our work aims to narrow the interest-to-adoption gap for decentralized social applications by providing a structured framework that helps activist communities form mental models for navigating the DSN landscape.~\cite{denardis_3_2022}.
To explore this space and understand its possibilities for community empowerment, we pose the following research questions:
\begin{itemize}
    \item (RQ1): How do DSNs distinguish themselves from centralized social networks and each other?
    \item (RQ2): How do the affordances of DSNs shape their social effects, particularly for activist communities? 
    \item (RQ3): How can activist communities---who are especially targeted by online censorship, harassment, erasure, and surveillance---leverage those technical distinctions to explore decentralized technologies that align with their goals and constraints?
\end{itemize}
In approaching these questions, our work connects the technical affordances of DSNs to the community interactions and formations they support. We contribute the following:
\begin{itemize}
    \item A conceptual framework for navigating the DSN landscape that defines four core community needs---minimal overhead, community building and reach, on- and offline safety, and operational sustainability---and links them to concrete platform affordances such as interface familiarity, discoverability, data privacy, and platform-level support (Section ~\ref{sec:framework}). 
    \item Applications of the framework to (1) evaluate and compare two DSNs, illustrating their sociotechnical tradeoffs; (2) characterize community configurations across the DSN landscape, exploring how technical affordances shape possibilities for collective action; and (3) two case studies grounded in real-world circumstances, demonstrating how the framework can support communities determine which platforms align with their concerns and constraints (Section ~\ref{sec:applying_framework}).
\end{itemize}

Our work contributes to broader threads of CSCW research that propose frameworks to address sociotechnical gaps ~\cite{ehsanChartingSociotechnicalGap2023}, study activist communities' use of social media \cite{wongJustDontQuite2024,irannejadbisafarSupportingYouthActivists2020,elmimouniExploringAlgorithmicResistance2025,ghoshalRoleSocialComputing2019, andrusDataActivismActivism2025, karizatAlgorithmicFolkTheories2021}, and characterize online communities within decentralized social networks \cite{hwangTrustFrictionNegotiating2025,zhang2024trouble}. By bridging these areas, we offer a framework that connects the technical affordances of DSNs to the needs of activist communities, thereby supporting the exploration and comparison of platforms. The framework also provides a foundation for future co-design of community-facing tools to support critical literacy and decision-making around DSNs. More broadly, we argue that questions of infrastructure are central to understanding how collective action is enabled, constrained, and sustained in digital environments.


\section{Background}

\subsection{Origins and Core Characteristics of Decentralized Social Networks}

\subsubsection{Origins of the Modern Social Networks and Emergence of Decentralization as an Alternative}
Current social networks can be characterized by a key structure and a set of core operations. The central structure is the social graph, which identifies the entities present in the network and describes their relationships. The core operations include identity management, authentication, data storage, content discovery, monetization, and content moderation.~\cite{boyd_social_2007, roscam_abbing_decentralised_2023, gillespie_custodians_2018}.
This contemporary social media paradigm is intimately tied to the ``corporatization'' of social interactions on the internet, driven by the rapid rise, diversification, consolidation, and influence of social network sites backed by private companies~\cite{boyd_social_2007, gillespie_politics_2010}.
To challenge this totalizing control of (online) social interactions by a few corporations, tech enthusiasts and proponents of a free and open Internet countered with Decentralized Social Networks~\cite{roscam_abbing_decentralised_2023, livitckaia_decentralised_2024}. From its inception, the decentralized social network ecosystem presented itself not only as a technical alternative but also as a deeply sociopolitical one, promoting a free speech and anti-monopoly ethos~\cite{denardis_3_2022}.
But despite their promises, decentralized social technologies have yet to face mainstream adoption~\cite{denardis_3_2022, frost-arnold_beyond_2024}. In addition to the inherent technical complexity of decentralized systems, the space of decentralized social networks is vast, comprising varying network types and an extensive set of differing protocols~\cite{balduf_looking_2024, datta2010decentralized, federation_federation_2024, la_cava_understanding_2021}. As a result, understanding the distinction between network types, protocols, and their implementations while navigating hundreds of decentralized social platforms becomes exceptionally challenging.
In this paper, we aim to provide a clear characterization of the decentralized social network landscape, especially for those whose identities and ambitions for collective action have made them targets of centralized (corporate) social media.

\subsubsection{Social Network Architecture Types}
Beyond impacting a system's capabilities and performance, social network architectures reify intentional social modes of organizations ~\cite{roscam_abbing_decentralised_2023}. On the one hand, social network architectures reflect technical considerations, including system availability, fault tolerance, and scalability. On the other hand, they reveal social concerns of ownership, power, and trust within the system~\cite{halpin_socio-technical_2023}. For instance, while a technical assessment of a system might ask ``can a single power outage cause the entire system to shut down?'', a sociotechnical assessment will ask ``can a single authority cause the entire system to shut down?''.
From a sociotechnical perspective, social network types can be categorized as \textit{centralized} and \textit{decentralized}.
Centralized social networks are defined by a single, trusted authority that manages all user data and activity~\cite{adhikari_vivisecting_2012}. Such a setup implies limited data privacy and security, and unilateral content moderation~\cite{gillespie_custodians_2018, barrigas_overview_2014}~\footnote{In practice, to efficiently manage their data, centralized social networks are supported by federated or distributed infrastructures.}.
In contrast, decentralized social networks, whether federated or distributed, lack a single source of trust. 
Federated social networks are defined by a collection of \textit{instances} (server nodes) that own and manage their users' data and activity, and that can interact with one another. Federated social networks are thus centralized at the instance level (i.e., semi-trusted) and decentralized at the network level, housing a set of instance-level authorities within an interconnected ecosystem. The ecosystem presents bespoke, instance-level content moderation, and offers a greater degree of freedom to network participants who can interact with different instances within a platform (\verb|Mastodon_x| to \verb|Mastodon_y|), as well as across platforms (\verb|Mastodon_x| to \verb|Pixelfed_x|)\footnote{Provided the platforms implement the same protocol; in this instance, the federated protocol ActivityPub.}. ~\cite{matney_twitters_2021, roscam_abbing_decentralised_2023}.
In a distributed social network, there is no central authority (i.e., the system is trustless). Each node manages its own storage and activity and can potentially connect to any other node. This generally offers greater user control over data, more granular content moderation, censorship resistance, and alternative avenues for monetization~\cite{steemit_steemit_nodate, manyverse_manyverse_2024, freight_deso_nodate, leitao_epidemic_2007}. However, distributed systems are also notoriously complex and challenging to design, maintain, and debug~\cite{jiang_survey_2016, naik_demystifying_2021}.

\subsubsection{Protocols: Blueprints of Decentralized Social Networks}
Key to the understanding of decentralized social networks is their underlying protocols. Protocols extend system architectures and represent the design patterns, instructions, and standards used to create client applications and platforms.
Protocols in the decentralized ecosystem are defined by three design principles: decentralization (i.e., no single authority), interoperability (i.e., interconnectedness between entities), and improved user control (i.e., user-level privacy, moderation, and content curation)~\cite{graber_applicationsblockchain-socialmd_2020, la_cava_understanding_2021}.

The manifestations of these design principles can be categorized into peer-to-peer (P2P), blockchain-based, and federated protocols. Peer-to-peer social network protocols like Secure Scuttlebutt (SSB) emphasize user control, data resilience, and privacy by enforcing a web-of-trust model, hosting user content on their machine, and enabling offline usage~\cite{scuttlebutt_scuttlebutt_nodate, manyverse_manyverse_2024, wei_exploring_2024}. Blockchain-based protocols like Farcaster use general-purpose blockchains (e.g., Ethereum) for core identity functions~\cite{steem_steem-whitepaper_2018}, while user-facing operations (e.g., posting, reacting, following) typically occur off-chain~\cite{farcaster_farcaster_2025}. Participants still generally need some blockchain literacy and a digital wallet, particularly for registration or monetization.
Carrying fewer design complexities than P2P and blockchain-based protocols, federated protocols like ActivityPub and the Authenticated Transfer Protocol (ATProto) support Mastodon and Bluesky, respectively---the two most popular decentralized platforms to date.
Protocols often implement hybrid architectures to balance data protection and performance considerations~\cite{datta2010decentralized, jeongNavigatingDecentralizedOnline2025}. For instance, Farcaster uses a federated network of Hubs at the data layer in addition to its blockchain-based identity layer~\cite{farcaster_farcaster_2025}.  

In their sociotechnical framework for characterizing protocols, Halpin, Brekke, and Issakides~\cite{halpin_socio-technical_2023} distinguish between a protocol's principles, technical properties, and social effects. Echoing Ackerman's definition of the ``socio-technical gap'' inherent in CSCW~\cite{ackermanIntellectualChallengeCSCW2000}, they argue \textit{principles} are design patterns for protocols that ``express the general assumptions of how to achieve certain technical \textit{properties} and social \textit{effects}'', and emphasize that protocol design principles do not necessarily lead to their intended social effects in practice and over time~\cite{halpin_socio-technical_2023}. In effect, architectural complexity exacerbates this distance, thereby challenging properties such as security and privacy~\cite{albertErrorAttackTolerance2000}. Building on this distinction, our work examines how technical properties interact with social effects by connecting the social media needs of activist communities with the affordances of decentralized protocol implementations (i.e., decentralized social networks).

\subsection{Activist Communities and Social Media}
Online communities have long used social media platforms to connect, collaborate, and advocate for shared causes and interests. They represent identity-based communities and practice-based collectives focused on areas such as open-source development, transformative works, and climate justice~\citep{jackson2020hashtag, fieslerMovingLandsOnline2020}. Such communities have used online spaces to cultivate identity, exchange knowledge, co-create content, and organize for political and social change~\citep{allredBeGayCrimes2021, payneBuildingSolidarityHostility2025, klassenBlackFuturePower2024, ghoshalRoleSocialComputing2019}.

In this work, we focus on \textit{action-capable collectives}---groups often referred to as activist communities, online collectives, or digital movements. These communities are not merely social or interest-based; they are organized, institutionalized, and oriented toward action. They typically exhibit three core characteristics: (i) a shared set of internal norms and rules; (ii) a collective identity that shapes both internal practices and public-facing activities; and (iii) an evolving internal differentiation, where leadership structures emerge over time to include organizing cores and influential activists, surrounded by networks of supporters and contributors~\citep{dolataMassesCrowdsCommunities2016}.

This structural differentiation leads to diverse governance models, ranging from hierarchical (top-down or ``feudal'') to fully horizontal, consensus-driven systems. Examples of action-capable collectives in the decentralized social network space include federated and peer-led communities like \texttt{blacksky}, \texttt{climatejustice.social}, and \texttt{spore.social}. While these groups span the spectrum of what might be called ``activist'' or ``online communities'', we adopt the term \textit{activist communities} throughout this paper to specifically denote action-capable collectives with a mission-driven focus.

Activist communities face a unique set of needs and constraints when using social media, such as maintaining security and autonomy, building trust, effectively mobilizing, scaling safely, and resisting surveillance or censorship~\cite{gonzalez-bailonNetworkedDiscontentAnatomy2016}. While the decentralized social media ecosystem offers promising alternatives to traditional, corporate-owned platforms, it remains vast, fragmented, and often difficult to navigate for newcomers. Recent studies have mapped the broader DSN landscape~\citep{jeongNavigatingDecentralizedOnline2025}, but these accounts are aimed at general audiences and do not center the specific needs of grassroots organizations or activist communities.

In response, this paper offers a community-oriented lens on DSNs, highlighting the platform affordances most relevant to activist communities and helping them make informed decisions about which implementations best align with their values, needs, and available resources. We introduce our framework next.


\section{A Framework to Connect Community Needs with Decentralized Platform Affordances}
\label{sec:framework}

In this section, we present our framework for systematically highlighting the trade-offs and promises of DSNs for activist communities. We outline the framework's synthesis process and describe its resulting dimensions.

Firstly, we distilled four community needs, which are informed by literature on activists' use of centralized social media and extended to encompass the unique considerations of DSNs: {\textit{minimal overhead}}, {\textit{community building and reach}}, {\textit{on- and offline safety}}, and {\textit{operational sustainability}} (\ref{sec:needs}).
Then, drawing from literature on DSN architectures and protocol design \cite{halpin_socio-technical_2023, jeongNavigatingDecentralizedOnline2025}, we extracted the core properties underpinning decentralized social network architectures to form a preliminary framework. We then iteratively refined the framework by adapting or combining affordances such as visibility, association, and meta-voicing from Guan et al.'s Web3 social media affordance framework~\cite{guanUsingAffordanceUnderstand2025}, and by integrating DSN documentation~\cite{mastodon_mastodon_2024, bluesky_blueskys_nodate, steemit_steemit_nodate, noauthor_a16zawesome-farcaster_2025} to contribute financial and hardware-related affordances. Lastly, we connected the distilled community needs to the redefined affordances of decentralized social networks (\ref{sec:needs_to_affordance}). These affordances constitute the 15 dimensions of the framework. 

For each dimension (e.g., data privacy), we provide examples of its implementation across different DSNs (e.g., granular privacy controls, identity verification mechanisms, encryption schemes). We also define features to characterize DSNs along these dimensions (e.g., data privacy is enabled by account anonymity/pseudonymity, content visibility controls, and identity verification) to facilitate cross-platform comparison. 

The framework's community needs and associated affordances are summarized in Table~\ref{tab:affordances}. The related features for each framework affordance are described in Appendix~\ref{appendix}. 

\subsection{Community Needs}
\label{sec:needs}
To distill activist community needs, we draw from and synthesize literature on activists' use of and experiences with (centralized) social media, adjusting these insights for the DSN ecosystem. 

Literature on the experiences and needs of activist communities on social media platforms cover a range of communities and social movements, including LGBTQ+ information access \cite{jonasBetterGoogleInformation2024}, Indigenous social movements \cite{lupienIndigenousMovementsCollective2020, duarteConnectedActivismIndigenous2017}, human rights activism \cite{elmimouniExploringAlgorithmicResistance2025}, climate justice movements \cite{lutherSocialMediaActivists2025, wongJustDontQuite2024}, and youth-led activism \cite{irannejadbisafarSupportingYouthActivists2020}. Across these different contexts, common themes emerge around communities' needs on social media platforms. All studies cite information access and community reach as a key need, especially in the face of censorship \cite{elmimouniExploringAlgorithmicResistance2025}, misinformation \cite{lutherSocialMediaActivists2025}, or physical disconnectedness \cite{jonasBetterGoogleInformation2024}. Many also highlight the importance of active community building, often centered around a particular identity \cite{jonasBetterGoogleInformation2024,duarteConnectedActivismIndigenous2017,lupienIndigenousMovementsCollective2020,wongJustDontQuite2024}. Physical and online safety is another common underlying need that manifests in communities' desire for anonymity \cite{wongJustDontQuite2024,jonasBetterGoogleInformation2024}, data ownership, or effective filtering of hate speech and harassment \cite{lutherSocialMediaActivists2025}.

Since these studies focus on needs arising from communities' use of centralized social media, we also draw from literature on the implementation and adoption of DSNs \cite{datta2010decentralized, halpin_socio-technical_2023,jeongNavigatingDecentralizedOnline2025} to understand unique needs that may emerge in this context. This literature addresses data management, resource requirements (e.g., for joining or maintaining a DSN instance), and openness (e.g., permissions or technical familiarity required to join the system). 

Synthesizing insights from activists' use of social media with the unique considerations of DSNs, we draw out four overarching community needs that guide the framework: (1) \textit{minimal overhead}: the ability for community members to engage in the platform with as little cognitive, technical, and financial cost as possible; (2) \textit{community building and reach}: the capacity to find, connect, and grow the target community; (3) \textit{on- and offline safety}: mechanisms that safeguard users from hate speech, harassment, targeting, or physical harm; and (4) \textit{operational sustainability}: the structures that support long-term viability and autonomy. 

\subsection{Connecting Community Needs to DSN Affordances}
\label{sec:needs_to_affordance}
We now connect the community needs defined above to the affordances of DSN platforms, resulting in the fifteen dimensions of the framework. To do so, we draw on platform-specific documentation and overviews of decentralized architectures and protocols \cite{halpin_socio-technical_2023,jeongNavigatingDecentralizedOnline2025}. 
Many of the proposed affordances are relevant to users of centralized social media platforms as well (e.g., customizability), but carry unique considerations in the decentralized context (e.g., the space of customization options is more extensive and variable across DSNs). For each affordance, we provide a brief description, explain how it might vary across DSNs, and define a set of features to characterize DSNs along that dimension.

As we provide a set of platform features that operationalize each affordance, we recognize that activist communities may prioritize these features differently depending on their specific needs. For instance, communities seeking to protect themselves from surveillance may consider data privacy features such as pseudonymity and post visibility controls to be critical, whereas communities focused on broad, cross-platform reach may place greater importance on interoperability features, such as cross-posting tools. Interface familiarity features may be deemed equally important by both communities. We believe that a qualitative assessment of the framework by members of activist communities can help surface similar nuances. We illustrate the ranking rationale using the case studies in Section~\ref{sec:case} and outline an evaluation plan in Section~\ref{sec:limitations}.

We aim for the framework to provide a mental model of the complex space of DSNs, a concrete analytical framework for evaluating and comparing specific platforms with respect to community needs, and a blueprint for developing community-oriented tools and guides for DSN exploration.

\subsubsection{Minimal overhead}
To meaningfully participate in decentralized spaces, activist communities should be able to join and engage without facing technical, financial, or cognitive barriers. DSNs differ in how easily new participants can get started, navigate interfaces, and maintain a presence over time. For example, DSNs vary in their degree of familiarity with mainstream platforms (\textit{interface familiarity}) and in whether they impose technical obligations, such as self-hosting (\textit{progressive hardware requirements}).

\medskip

\noindent\textbf{Account creation accessibility.}
It is the affordance that enables visitors to become registered participants and is foundational to their appreciation of the value of a digital platform.
An overly complex account creation experience can leave participants confused and frustrated, ultimately reducing their engagement and increasing the likelihood of abandonment~\cite{venkateshConsumerAcceptanceUse2012}.
Account creation accessibility is defined by zero-verification sign-up (i.e., identity verification without email or phone number); multimodal access via mobile or web operating systems and online or offline (SMS, Bluetooth); few setup steps; internationalization and visual accessibility; and migration and bridge tools.

\medskip

\begin{table}[htbp]
\centering
\caption{Overview of the framework's affordance dimensions, organized by the four community needs they address, with operational definitions for each affordance.}
\label{tab:affordances}
\small
\begin{tabular}{p{2cm}p{3.5cm}p{7.5cm}}
\toprule
\textbf{Community Need} & \textbf{Affordance} & \textbf{Definition} \\
\midrule

\multirow{5}{2cm}{Minimal overhead} 
& Account creation accessibility
& The ease with which visitors can become registered, participating users. \\
\cmidrule(lr){2-3}

& Interface familiarity
& The extent to which users can navigate the interface using prior experience, without instruction or mental translation. \\
\cmidrule(lr){2-3}

& Smart defaults
& The extent to which participants can engage immediately, without configuration or decision-making, due to reasonable system-set defaults. \\
\cmidrule(lr){2-3}

& Hardware accessibility
& The extent to which the platform is usable on low-resource devices. \\
\cmidrule(lr){2-3}

& Financial accessibility
& The extent to which users can participate meaningfully without cost. \\
\midrule

\multirow{4}{2cm}{Community Building and Reach}
& Discoverability
& The extent to which users can find relevant content, communities, or other users. \\
\cmidrule(lr){2-3}

& Interoperability
& The platform's ability to exchange content, identity, or data with other systems or platforms. \\
\cmidrule(lr){2-3}

& Customizability
& The extent to which users can modify content, interface, and platform behavior beyond defaults. \\
\cmidrule(lr){2-3}

& Interactivity
& The extent to which users can form relationships with others and respond directly to content. \\
\midrule

\multirow{3}{2cm}{On- and offline safety}
& Data privacy
& The extent to which users can control the visibility of their profiles, relationships, and actions. \\
\cmidrule(lr){2-3}

& Data ownership and control
& The extent to which users can determine how their data is stored, secured, migrated, and deleted. \\
\cmidrule(lr){2-3}

& Moderation
& The extent to which a platform affords users the ability to restrict certain content and accounts. \\
\midrule

\multirow{3}{2cm}{Operational sustainability}
& Governance structures
& The extent to which decision-making authority and access rights are assigned among users, administrators, and developers. \\
\cmidrule(lr){2-3}

& Platform-level support
& The extent to which users can rely on consistent assistance and maintenance to keep the platform functional and secure. \\
\cmidrule(lr){2-3}

& Transparency
& The extent to which platform operators openly communicate about system infrastructure, policies, and established processes. \\

\bottomrule
\end{tabular}
\end{table}

\noindent\textbf{Interface familiarity.}
This affordance refers to users' ability to understand and use an interface based on prior experience with similar systems, without explicit instruction or mental translation~\cite{baj-rogowskaExploringUsabilityUser2023, nielsenUsabilityEngineering1994}. For DSNs, we highlight two core features of interface familiarity: (1) established visual, conceptual, and interaction conventions; and (2) minimal jargon, recognizing that users will use knowledge from past social media user experiences when adopting new platforms ~\cite{fieslerMovingLandsOnline2020}.

\medskip

\noindent\textbf{Smart defaults.}
Inspired by prior work on usability heuristics~\cite{nielsenUsabilityEngineering1994, nielsenEnhancingExplanatoryPower1994}, we define smart defaults as the extent to which users can use the platform immediately, without configuration or decision-making, because the platform has made reasonable default configuration choices. Depending on the DSN, hidden or unintended default settings can range from degrading activists' first impressions of a platform to inadvertently exposing their identities. The features that constitute this affordance include extensive privacy settings, available default instances, domains, or nodes, and a pre-populated content feed.

\medskip

\noindent\textbf{Hardware accessibility.}
Activist communities often operate with limited resources, making social media a powerful and cost-effective tool for gaining visibility, expanding networks, and organizing large-scale actions~\cite{earlDigitallyEnabledSocial2011}. While mainstream platforms are typically free to use but monetize user data, some decentralized social networks (DSNs) may introduce costs, such as storage fees, backup services, or self-hosting requirements. 
Hardware accessibility refers to the extent to which the platform is usable on low-resource devices without requiring high-end hardware to participate. Features include low initial storage and compute footprint, cloud-hosted infrastructure, and the availability and quality of related documentation. 

\medskip

\noindent\textbf{Financial accessibility.}
Additionally, some DSNs offer paid tiers with different functionality. For example, Farcaster offers higher character limits and more image uploads in its ``Pro'' tier, available through subscription, whereas all of Bluesky's features are free for users at the time of writing. We refer to the extent to which users can participate meaningfully at no cost, regardless of their ability to pay for premium features or for optional paid features that provide advanced capabilities, as the financial accessibility affordance. It is operationalized as the free-to-paid feature ratio, defined as the number of free features divided by the number of premium features (the higher the better). This determines whether activists can access and sustain the platform at little to no financial cost.

\subsubsection{Community building and reach}
To cultivate their community, interact with other communities, garner media attention, build political pressure, or quickly mobilize, activists must identify their target audience and effectively disseminate educational content or mobilization messages. DSNs introduce unique affordances for community building and reach, which we highlight in this section: for example, the scale and speed of content discovery and propagation vary across protocols (\textit{discoverability}), and some protocols offer support for cross-platform posting and browsing (\textit{interoperability}).  

\medskip

\noindent\textbf{Discoverability.}
Extending Guan et al.'s searchability affordance (i.e., the ability of users to discover and access content from others \cite{guanUsingAffordanceUnderstand2025}) discoverability refers to how easily users can find relevant content across a platform, particularly through features such as search functionality, visibility mechanisms that help users better control the content they wish to see, access modes that determine whether content can be viewed by non-authenticated users or users outside of the poster's social graph, and content synchronization (i.e., the speed and extent to which content propagates through the network). Peer-to-peer networks typically offer limited, local discoverability, while many blockchain-based platforms promote broader sharing and discovery. Federated networks fall somewhere in between.

\medskip

\noindent\textbf{Interoperability}
Interoperability is the ability of a platform to exchange content, identity, or data seamlessly with other systems or platforms. For activist communities, interoperability offers significant advantages, including broader reach, easier coordination across platforms, and reduced platform lock-in. This affordance includes support for open protocols---such as ActivityPub, AT Protocol, and Diaspora*---which enable users to post within a platform and across compatible platforms. Additionally, third-party tools such as Mastogram, Publer, Croissant, and OpenVibe enhance interoperability by enabling cross-platform posting and browsing, even when platforms use different protocols.

\medskip

\noindent\textbf{Customizability.}
Customizability refers to the range of options a platform provides for modifying content, the user interface, and user actions beyond the platform's defaults. We frame this as a usability-related affordance to emphasize the varying levels of personalization available across decentralized social networks (DSNs). Its features include support for multiple content formats, such as text, audio, and video; content updates; and profile and user interface customization. Platforms such as Misskey and Pleroma offer extensive customizability, allowing users to tailor timelines, themes, and feature sets. In contrast, platforms such as Nostr often prioritize privacy, security, and offline-first functionality and may limit or omit user-facing customization features.

\medskip

\noindent\textbf{Interactivity.}
Combining the definitions of association and meta-voicing affordances~\cite {guanUsingAffordanceUnderstand2025}, we define interactivity as the extent to which a platform enables users to form relationships with other users and to provide direct responses to content. Diverse association mechanisms and core content-interaction features characterize this affordance, namely, following, direct and group messaging, liking, commenting, and reposting. The importance of specific features may vary with community objectives (e.g., confidential coordination prioritizing direct messages versus wider-reach coordination preferring channels and group chats).

\subsubsection{On- and offline safety}
Activists facing doxxing, threats, and other forms of violence---which may extend beyond online spaces---require greater privacy and ownership over their profile, content, and interaction data.
DSNs vary in the degree to which users can manage the visibility, portability, and permanence of their digital footprints. For example, some platforms support pseudonymity and identity verification mechanisms (\textit{data privacy}), whereas others allow communities to self-host, migrate, and delete their own data (\textit{data ownership and control}).

\medskip

\noindent\textbf{Data privacy.}
Adapting the visibility affordance (i.e., ability to individually determine what content linked to one's persona is visible to others ~\cite{guanUsingAffordanceUnderstand2025}), this dimension refers to the extent to network participants can control the visibility of their profiles (e.g., username, name, email, profile photo), relationships (e.g., followers, friends), and actions (e.g., posts, likes, comments, DMs). Activists and members of communities of interest often prioritize data privacy to protect themselves from harassment, surveillance, and other threats. These concerns are vital for both mental well-being and physical safety. At the same time, trusted identity verification mechanisms are needed to prevent spam and bot activity. We associate data privacy with account anonymity or pseudonymity (i.e., the ability to maintain a consistent identity without revealing real-world identifying information); end-to-end message encryption; content visibility controls; and identity verification tools.

\medskip

\noindent\textbf{Data ownership and control.}
This affordance represents users' or server administrators' ability to determine how their data is stored, secured, migrated, and deleted. Users should understand whether their members' profile, content, and activity data are stored on individual devices (peer-to-peer), on a shared server instance (federated), in a platform-level data structure such as a public ledger (blockchain), or through a combination of these methods (e.g., ATProto). Some platforms may offer data backup options, while others may not. Furthermore, communities should consider whether their data can be transferred to other platforms or groups. The features we consider to characterize data ownership and control include self-hosting and storage options, data download, data migration, and data deletion.

\medskip

\noindent\textbf{Moderation.}
Moderation is the extent to which a platform affords users the ability to restrict certain content and accounts.
Centralized platforms typically rely on top-down moderation administered by the platform itself, with limited user involvement. In contrast, DSNs often support a combination of platform-level and community-driven moderation. For example, Mastodon allows instance administrators and moderators to enforce local rules, while also giving users tools to block, mute, and report content. Bluesky offers composable moderation services that can be mixed and matched across clients, and Nostr relies heavily on user-controlled blocking and third-party relays for moderation, with minimal central enforcement. To evaluate moderation affordances, we consider three features:  the available safety tools (e.g., reporting, blocking, muting, banning) for activist communities to protect themselves from harassment or infiltration; reporting mechanisms (e.g., manual reporting, automated flagging) that can support community safety; and rule-sharing mechanisms, helpful for collective defense (e.g., blocklists, mutelists).

\subsubsection{Operational sustainability}
Activist communities often rely on volunteer labor and limited resources, making operational sustainability a core concern. DSNs vary in their support for long-term maintenance, their distribution of decision-making, and their visibility of internal processes. For example, platforms offer different decision-making structures (\textit{governance structures}), feedback channels, and formal support (\textit{platform-level support}), as well as visibility into protocol development and policy decisions (\textit{transparency}). 

\medskip

\noindent\textbf{Governance structures.}
How authority and access are assigned within a platform determines the degree of control held by different members of the community. 
We define the governance structures affordance as the way a platform facilitates or assigns decision-making authority, access rights, and policies within and across its architecture layers. Whether roles such as moderator, instance administrator, and ordinary user are concentrated among a few or distributed broadly across an online community's membership shapes the platform’s overall governance structure. Activist communities, in particular, may prefer governance models that align with their values---such as transparency, accountability, or collective decision-making---and reflect their available time and resources. To provide greater visibility into the layers of decision-making, we identify configurable decision-making mechanisms (e.g., support for proposal creation and voting) and instance-level and platform-wide privileges that determine a network participant's ability to propose, create, and decide on policies, guidelines, and features relevant to their community.

\medskip

\noindent\textbf{Platform-level support.}
While autonomy is essential for enhancing a community’s social experience on decentralized platforms, maintaining that autonomy often requires a baseline level of technical expertise. At the same time, safeguarding users against illegal content and harmful practices requires some platform-level support mechanisms. As such, even in decentralized systems, there is a need for reliable system maintenance and clearly defined support and feedback channels to address complaints, disputes, appeals, and general user feedback. For example, platforms such as Bluesky provide moderation services and reporting tools at the protocol and application levels, enabling users to raise concerns or flag content. In contrast, systems like Nostr may lack formal support structures entirely, relying instead on informal, community-driven responses. Therefore, we define platform-level support as the ability to rely on consistent, accessible assistance and maintenance from the platform's core team or community, ensuring the platform remains functional, secure, and responsive to user needs over time. Its characteristic features include maintenance resources and communication channels.

\medskip

\noindent\textbf{Transparency.}
Transparency refers to the degree of openness with which decentralized platform operators communicate about application features, platform-level policies (e.g., Terms of Service), and established processes for handling appeals and complaints. When information about a platform’s internal governance is consistently available and publicly accessible, it serves as a strong indicator of the level of safety and accountability that communities can expect when adopting the platform. To better capture the different facets of this quality, we distinguish between available open-source code, open APIs, organizational structure documentation, and policy and process decisions documentation.
For example, federated platforms such as Pleroma or Misskey offer open-source code, accessible governance documentation, and community-involved development. Bluesky, on the other hand, provides protocol-level transparency through its open ATProto documentation, while continuing to develop its community relations and communication practices. In contrast, platforms such as Nostr rely on minimal centralized communication.


\section{Applying the Framework}
\label{sec:applying_framework}
This section explores how the framework's affordances provide a structure for characterizing and comparing specific applications and contribute to a broader understanding of the types of community formations enabled across the DSNs. We begin by applying the framework to two representative platforms---Mastodon and Bluesky--- to show how their technical affordances shape the social configurations they enable (\ref{sec:pc}). We then draw on related work to understand broader patterns in community organization across different categories of DSNs (\ref{sec:comm}). Finally, we present two speculative case studies that illustrate how the framework might guide activist communities navigating high-stakes contexts---such as journalists responding to internet shutdowns (\ref{sec:case}). Our framework facilitates understanding the DSN landscape through the lens of community-oriented affordances, offering a mental model that supports sociotechnical analysis of both platforms and the communities they enable.

\subsection{Platform Cards: An Application-level Illustration of the Framework}
\label{sec:pc}

\begin{figure}[!htbp]
    \centering
    \includegraphics[width=.93\textwidth]{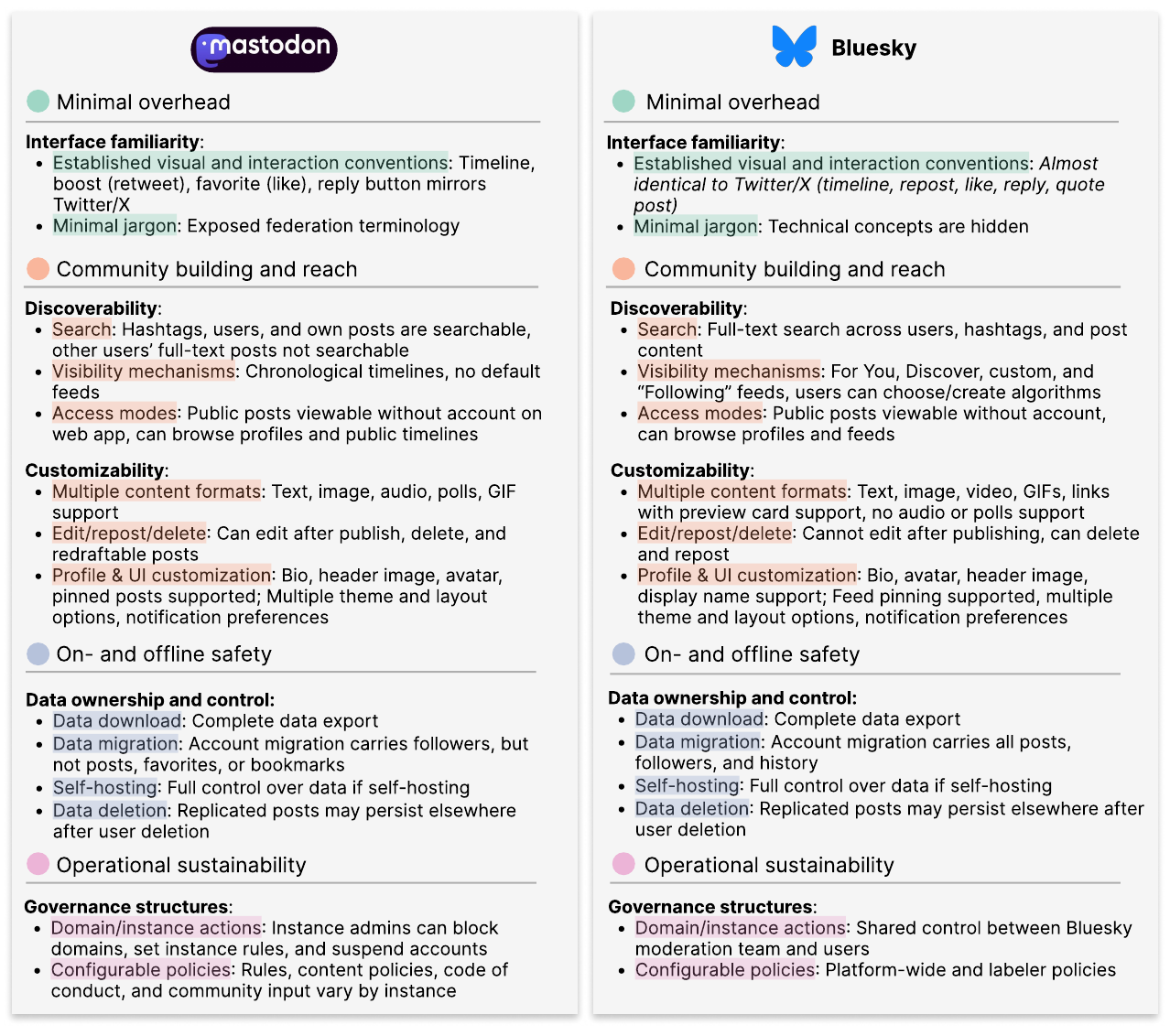}
    \Description{Decentralized Platform Cards}
    \vspace{-0.6em}
    \caption{\footnotesize Platform cards applying a subset of the framework to Mastodon and Bluesky. Each card summarizes how the DSN implements a subset of key affordances across four activist community needs, illustrating their sociotechnical tradeoffs.  }
    \label{fig:profiles}
\end{figure}

To ground the framework in the current DSN landscape, we preliminarily apply it to two of the most widely adopted decentralized platforms by total user count at the time of writing: Mastodon~\cite{mastodon_mastodon_2024} and Bluesky~\cite{BlueskyDocumentationBluesky}.
Figure~\ref{fig:profiles} presents this analysis as ``platform cards'' that summarize key features of each application, showing a representative subset of affordances (5 of 15) for readability. We note that in their current form, the platform cards serve as illustrative summaries rather than activist-facing decision-making tools. We envision them as a foundation for future work, in collaboration with activist communities, to co-design accessible guides and tools that directly support capacity-building and decision-making. Much like a protocol serves as a schema for implementing applications, our framework provides a blueprint for co-creating documentation and tooling to help action-capable (activist) collectives navigate and understand the DSN ecosystem.

We find that an initial application highlights how the framework reveals nuances in platforms' technical affordances, even when platforms may appear similar. For example, both Bluesky and Mastodon are decentralized microblogging platforms, but Bluesky offers lower \textit{overhead} than Mastodon, with an interface almost identical to Twitter/X, an absence of platform jargon (e.g., no explicit mention of DIDs or personal data servers), and a set of suggested accounts to follow during onboarding. Each platform also differs in its support for \textit{community building and reach}: Mastodon enables text, images, audio, polls, and GIFs, whereas Bluesky does not support polls at the time of writing. In contrast, Bluesky enables greater discoverability and content creation, allowing users to create and select multiple feeds.

Regarding \textit{on- and offline safety}, Mastodon offers more content visibility controls for posts than Bluesky, but fewer data controls, because account migrations across instances are lossy, whereas account migrations across personal data servers preserve all of the account's posts, followers, and history. Finally, we also observe tradeoffs for \textit{operational sustainability}. Mastodon grants each instance significant autonomy over its rules, moderation policies, and technical configurations, with instance administrators holding broad control while ordinary users have limited rights. Bluesky, by contrast, imposes a shared baseline of rules and policies but adopts a more modular approach that separates hosting, identity, and moderation into independent layers, allowing users to choose which services to own or rely on beyond the platform defaults.

We note that this analysis is preliminary and intended to demonstrate the framework's potential rather than offer definitive platform assessments. Still, these contrasts underscore how no single platform fully satisfies all community needs; instead, each foregrounds certain affordances while limiting others. The framework makes these trade-offs legible by enabling platform analysis in relation to community needs. These comparisons also point to broader patterns in how platform architectures give rise to distinct social and governance arrangements, which are analyzed further in the following section on community configurations across the decentralized social universe.

\subsection{Community Configurations in the Decentralized Social Universe}
\label{sec:comm}

Here, we extend the characterization of two exemplar platforms (Section \ref{sec:pc}) into a broader analysis examining how distinct technical architectures give rise to different community configurations across the DSN landscape. We use the term {\textit{community configuration} to describe the recurring social formations that emerge across platforms, including different arrangements of user roles, authority, and collective behavior. We analyze how platform affordances have already given rise to different community configurations, highlighting the sociotechnical conditions that may make certain types of activist communities resilient or fragile in decentralized environments. Our analysis extends prior technical overviews of the DSN landscape \cite{halpin_socio-technical_2023,jeongNavigatingDecentralizedOnline2025} with our framework to characterize current community configurations---and their implications for collective action---across three DSN infrastructures: peer-to-peer and blockchain-based platforms, the fediverse, and the ATProto ecosystem. 

\subsubsection{Peer-to-Peer and Blockchain-based: Users-as-Moderators, and Token Holders}
Due to the inherent complexity of distributed network architectures, P2P social networks continually face a trade-off between system performance and user experience. These platforms afford high levels of \textit{data ownership} and local \textit{governance structures}, but limited \textit{account creation accessibility}, \textit{discoverability}, and \textit{interoperability}~\cite{p2p_aether_2025, manyverse_manyverse_2024, briar_briar_2025}.
Because of these high entry costs, the size and homogeneity of the online communities that emerge on P2P platforms tend to be relatively small. As a result, P2P social network users are typically tech enthusiasts who are drawn either to the technical promises of these technologies or to the belief that they are the answer to the sociopolitical project of (digital) citizen freedom~\cite{denardis_3_2022}. Platforms like Briar, for example, support highly resilient and privacy-preserving communication among tightly knit groups, with strong affordances for on- and offline safety but minimal support for community reach or operational sustainability. The community configurations that emerge, then, tend to be self-governing and secure, but largely closed and low-visibility.

For blockchain-based social networks, their financialized nature is inextricably linked to cryptocurrencies, regardless of whether the cryptocurrencies are directly pegged to fiat currencies such as the USD~\cite{jang_user_2022, xavier_evidence-based_2024, sharma_future_2024}.
Consequently, many communities in blockchain-based social networks are interested in decentralized finance (DeFi)~\cite{sharma_future_2024} and do not generally engage in social-justice-driven collective action. 
However, committed to the participatory ideal of citizen power ~\cite{arnstein_ladder_1969}, Decentralized Autonomous Organizations (DAOs) are Web3 (i.e., blockchain-related) communities where members engage in collective decision-making, typically carried out through a series of proposals where they vote on organizational events using governance tokens that indicate relative influence within the DAO. 
The governance structure in the DAO is algorithmic, as rules of engagement are encoded in software called smart contracts~\cite{axelsen_you_2024}. Socially-oriented DAOs include in-person interactions~\cite{cabin_cabin_nodate}, encourage more deliberative empowerment~\cite{soleimanof_community_2024}, and display higher levels participation ~\cite{sharma_future_2024}. 
Nevertheless, DAO voting can be subject to governance abuses, particularly systemic bribery, proposal bundling, and the potential corruption of identity systems used in DAO voting, reflecting governance patterns present across Web3 communities \cite{austgen_dao_2023, chenDecentralizedWeb3NonFungible2025}.

Through the lens of the framework, we see that P2P and blockchain-based platforms generally produce community configurations that emphasize autonomy and control, but often limit access, reach, and scalability. Activist communities that might be resilient here include high-trust collectives operating with minimal infrastructure, as well as well-resourced groups capable of managing technical complexity and engaging in governance-heavy processes.

\subsubsection{Fediverse: Users and Server Administrators/Moderators}
Communities on platforms built atop federated protocols such as ActivityPub and diaspora inhabit a loosely connected constellation of independently governed servers---collectively referred to as the fediverse.
Regardless of platform (e.g., Mastodon, Pixelfed, Pleroma, diaspora*, or Lemmy), fediverse participants generally fall into one of three potentially overlapping roles: administrators, who set up and maintain independent server instances; moderators, who enforce community rules; and users, who interact within their instance and with users of other instances.
This instance-based design supports \textit{governance structures} with significant autonomy within the network/platform, as each server instance sets its own community guidelines, moderation policies, and technical configurations. The fediverse affords strong instance-level \textit{moderation} and \textit{transparency} but introduces hierarchical \textit{governance structures} at the server-instance level. Server admins can approve new users, manage infrastructure, moderate content, and block interaction from other instances~\cite{he_flocking_2023, rozenshteinModeratingFediverseContent2022}. Server users, meanwhile, can join an instance by agreeing to its community guidelines, interact with other instances and platforms based on their server's configurations~\cite{bustamante_governance_2023}, and migrate to other instances (though this may entail data loss)~\cite{tosch_privacy_2024}.
Because administrative and user privileges are strictly separated at the instance level, enabling collaborative leadership or participatory governance depends heavily on the instance admin's willingness, who controls both infrastructure and moderation~\cite{kleppmann_bluesky_2024}. 

Centralization dynamics can also manifest at the network level. For example, the small number of Mastodon instances that host large user bases and popular posts drives user- and content-driven pressure toward centralization, while lower costs and simpler setup at a few hosting providers create infrastructure-driven pressure toward centralization~\cite{raman_challenges_2019}.
Despite these structural and implicit constraints, the fediverse supports a wide range of users and community types. While many users are developers and tech enthusiasts, communities of the fediverse reflect a larger user base and a more diverse range of positionalites and interests compared to P2P platforms. Many fediverse communities interact in instances explicitly dedicated to social justice and activism. 
Communities on the fediverse have also engaged in collective action~\cite{black__2024}, drawing on local \textit{moderation} affordances to block alt-right instances and platforms that promote hate speech and misinformation~\cite{fence_garden_2025}, or overly influence the entire network due to the sheer size of their server instances~\cite{meta_introducing_2023}. At the same time, such community protection actions are susceptible to adverse effects, creating conflict within communities and affecting marginalized users~\cite{melderBlocklistBoundaryTensions2025}. Moreover, action-capable collectives in the fediverse still grapple with persistent issues of whiteness (reflecting the demographic biases of early adopters) and a lack of deep engagement with racial justice~\cite{kiam_blackness_2023, hendrix_whiteness_2022}. 

Overall, federated architectures enable community configurations that balance local governance structures with wide community reach, supporting a mix of public-facing communication and values-aligned enclaves. The fediverse is well-suited to support activist groups seeking both autonomy and collaboration, but their viability depends on instance norms and admins. 

\subsubsection{Emerging Communities of the AT Protocol Networks}
The Authenticated Transfer Protocol (ATProto)---and Bluesky, its flagship platform---was designed to address functional limitations of the fediverse and P2P protocols, particularly user-level control and content discoverability. However, some of its key infrastructure components remain centralized to deliver an improved user experience: direct messages, for instance, are managed by a centralized service run by Bluesky's platform operators~\cite{kleppmann_bluesky_2024}.

Still, because it decouples components of its architectural design (e.g., identity, content feeds, data hosting, and moderation services), ATProto achieves greater \textit{customizability}, enabling participants to reconfigure core aspects of their experience. This design principle is reflected in two of Bluesky's main features: ``algorithmic choice,'' in which network participants select feeds from a marketplace, and ``composable moderation,'' in which participants create and publish moderation services for custom content filtering.
Together, these affordances support more flexible community configurations, in which ordinary users, developers, and moderators can assume shifting, interconnected roles. As a result, many individual users and online communities have taken advantage of Bluesky and ATProto's relative openness to build user-driven tools and community spaces, releasing labelers, feed generators, and custom application clients, or building entirely new social media platforms~\cite{perez_how_2024, fishttp_fishttpawesome-bluesky_2024}. Furthermore, communities that have been overly policed by centralized social networks have increasingly organized on Bluesky based on shared identities and interests. A notable example is {\itshape blacksky}\footnote{https://www.blackskyweb.xyz/}, which leverages ATProto to build a suite of tools for curating Black-centered content~\cite{balduf_looking_2024}. That said, while composable moderation enables activist communities to limit harmful content, Bluesky's baseline moderation layer is enforced across the network, leaving platform-level censorship a remaining reality for activist communities.

Bluesky and ATProto are relatively new, so their recent surge into the mainstream warrants careful future examination of their social impacts in addition to large-scale analyses~\cite{balduf_looking_2024, baldufBootstrappingSocialNetworks2025}. Our framework suggests, for instance, that Bluesky PBC\footnote{Public Benefit Corporation} faces limited \textit{transparency} and that the network presents widely varying \textit{progressive hardware requirements} for self-hosting. For example, running a personal data server requires only a small amount of storage for a small number of users, whereas hosting a relay requires substantial computing resources to crawl the entire network~\cite{HowDecentralizedBluesky}.

Overall, our analysis demonstrates how decentralized architectures give rise to distinct community configurations. Platforms built on P2P, blockchain-based, federated, and hybrid protocols each foreground different affordances and constrain others: P2P platforms prioritize privacy and security, but may limit discoverability and usability; federated platforms offer instance-level governance and interoperability, but rely heavily on instance admins and moderators; and composable protocols like ATProto support customizability but rely on evolving governance structures. 
Applying our framework helps understand these broader tradeoffs in the landscape of DSNs that inform how different platforms may or may not serve the needs of diverse activist communities.

\subsection{Case Studies}
\label{sec:case}
Here, we examine how the framework's affordances might help activist communities navigate the DSN landscape in practice. We describe two real-world scenarios in which activist communities face infrastructural and sociopolitical constraints that might motivate the adoption of DSNs: (1) political organizing and journalism during internet shutdowns, and (2) worker organizing under the threat of corporate surveillance. These case studies illustrate how the framework can support exploration of the decentralized ecosystem, helping communities align collective goals with platforms' technical affordances.

\subsubsection{Case 1: Activism in the Face of Internet Shutdowns}
Journalists and political activists are among the most targeted individuals within activist communities worldwide. These groups routinely face heightened risks of surveillance, censorship, platform shutdowns, and threats to their physical safety. Prior work has documented the digital vulnerabilities of these communities, including government surveillance, doxxing, and repression through digital infrastructure~\cite{earlDigitalRepressionSocial2022, hanPressProtectHelpingJournalists2024}.

The year 2023 was particularly alarming for digital repression. According to the digital rights organization AccessNow\footnote{\url{https://www.accessnow.org/}} and the \#KeepItOn coalition, there were 283 internet shutdowns across 39 countries---a 41\% increase from the previous year---marking the highest number of recorded shutdowns since monitoring began in 2016~\cite{tackett_internet_2024}. Notably, two of these shutdowns were externally imposed: Israel on Palestine and Russia on Ukraine. Many others occurred during periods of protest, political unrest, or armed conflict, while a smaller but emerging number were triggered by natural disasters~\cite{rightswatch_libya_2024}. 

These systematic shutdowns reflect a growing trend of governments increasingly leveraging their control over centralized infrastructure to stifle dissent, obstruct communication, and destabilize activist networks. Centralized social media platforms have proven especially vulnerable. First, they provide states with vast surveillance capabilities---enabled by metadata collection, account tracking, and content filtering~\cite{dean_myanmar_2017, nurik_facebook_2022}. Second, they are susceptible to state intervention and blocking, making them unreliable during politically sensitive moments~\cite{solomon_social_2024, xynou_grindr_2023}.

In response, activists have developed creative strategies to counteract state-imposed limitations. These include evasive communication, encrypted messaging apps such as Signal, mesh networks such as Bridgefy during protests, and circumvention tools such as VPNs and Tor~\cite{bhatiaProtestsInternetShutdowns2023, elmimouniExploringAlgorithmicResistance2025}. Still, these ad hoc solutions often fall short in situations of sustained repression or complete network blackouts. In such scenarios, decentralized social networking (DSN) platforms offer promising alternatives by supporting both digital sovereignty and operational continuity, even in compromised network environments.

Drawing from the community needs identified in Section~\ref{sec:needs}, we identify the following priority requirements for journalists and political activists operating under these conditions. Given the acute risk of surveillance and infrastructure disruption this community faces, safety and minimal overhead emerge as the most pressing needs, taking precedence over broader mobilization or reach objectives that other activist communities might prioritize. These needs map to the following platform affordances discussed in Section~\ref{sec:needs_to_affordance}:

\begin{itemize}
    \item \textbf{On- and offline safety:} Platforms should provide data privacy and data ownership and control by ensuring anonymity or pseudonymity, with the option to verify identities, as well as the possibility for their participants to store, migrate, or delete their data to protect themselves against surveillance and exposure.
    \item \textbf{Minimal overhead:} A simple onboarding experience is essential, especially for network participants under stress or with limited technical expertise. This can be achieved through accessible account creation via multimodal access (web, mobile, SMS, on- and offline), smart default privacy settings, and a familiar interface with little to no technical jargon. Given the immediate constraints, hardware resource efficiency is also key; platforms should be low-bandwidth, low-storage, and low-power.
\end{itemize}

Within these affordances, not all features carry equal weight for this community. Pseudonymity and offline functionality are indispensable: their absence would either expose members to direct harm or render the platform unusable during a shutdown. Data portability and end-to-end encryption matter substantially, given their importance for long-term operational security, though the community could likely adapt if these features were imperfect or partially available. Features such as cross-posting tools or rich media support, while potentially valuable in other contexts, are largely peripheral to this community, as they do not directly address the immediate threats of surveillance or loss of connectivity.

Under these criteria, a peer-to-peer (P2P) application such as Briar~\cite{briar_briar_2025} emerges as a strong candidate. Designed explicitly for journalists and activists, Briar operates without central servers and can synchronize messages over Bluetooth, Wi-Fi, or the internet, thereby remaining functional during internet shutdowns. However, its onboarding process is somewhat technical, which is a meaningful trade-off for the minimal overhead it affords and may pose challenges for less-experienced users.

Although Briar is not optimized for mass broadcasting, it can be paired with other tools---whether decentralized (e.g., Mastodon, PeerTube) or centralized (e.g., Twitter, YouTube)---to maintain a layered communication strategy. In this hybrid model, secure platforms are used for internal planning, while public platforms serve broader outreach and visibility goals. 

This case study illustrates how the framework might help communities evaluate platforms not only on ideological or technical grounds, but on how well they meet immediate needs for safety and resilience. 

\subsubsection{Case 2: U.S. Union Activity and the Need for Secure Digital Organizing.}

Despite the relatively unchanged union membership rate from 2022 to 2023~\cite{bls_union_2024}, 2023 witnessed a marked surge in union activity in the United States~\cite{bls_work_2024}. This rise in organizing efforts captured significant media attention and helped reintroduce the concept of worker power into the public imagination. It also coincided with a broader context of dissatisfaction among U.S. workers---only about half reported being very satisfied with their jobs overall, and just one-third expressed satisfaction with their income or opportunities for advancement~\cite{parker_how_2023}.

While worker organizing gained traction, a volatile policy landscape added further complexity. The U.S. Department of Labor's July 2024 salary threshold expansion, followed by its repeal in November 2024, generated considerable uncertainty and resentment, particularly among professional and administrative employees~\cite{dol_biden-harris_2024, advocacy_federal_2024}. At the same time, reports of union suppression proliferated: Starbucks faced hundreds of unfair labor practice charges filed by the National Labor Relations Board (NLRB)~\cite{reutersCourtRejectsStarbucks2024}, and Amazon was accused of illegal retaliation against warehouse workers organizing for better conditions~\cite{usaNLRBAmazon}.

However, challenges remain. National labor organizations risk becoming more conservative in their political alliances under the current U.S. administration. Moreover, increasing collaboration between government officials and CEOs of major social media companies has raised concerns about the expansion of online surveillance, particularly in monitoring labor organizing and protest planning~\cite{york_silicon_2021, klonick_new_2018}.

In this environment, workers seeking to organize and advocate for their economic futures must look toward alternative digital infrastructures. These platforms must enable secure coordination, resource sharing, meeting planning, and long-term community building~\cite{boag_tech_2022, yaoTogetherAloneAtomization2021}. Unlike the previous case, where immediate physical safety and offline functionality were paramount, this community's threat model is shaped by sustained surveillance and institutional retaliation, making community building and reach equally important alongside safety. Applying our framework, we identify the following key needs and corresponding platform affordances:

\begin{itemize}
    \item \textbf{Community building and reach:} Platforms should be searchable, provide multiple content formats, and facilitate communication across instances to allow participants to securely interact and share resources.
    \item \textbf{On- and offline safety:} Platforms should ensure anonymity or pseudonymity, with the option to verify identities, the possibility to migrate instances without data loss, block, mute, and filter out bad actors, safeguarding users from surveillance, doxxing, and retaliatory actions.
    \item \textbf{Operational sustainability:} Platforms should support community-level governance through flexible policies and robust autonomy in making moderation decisions.
\end{itemize}

Within these affordances, pseudonymity and instance-level moderation controls are indispensable, as their absence could expose members to retaliation or undermine self-governance. Data portability and cross-instance communication matter substantially, given their importance for sustaining organizing networks, though the community could likely adapt if these features were imperfect or partially available. Searchability and multi-format content support offer meaningful but non-essential value, supporting day-to-day organizing without addressing a core vulnerability. Offline mesh networking, by contrast, which was indispensable in the previous case, becomes largely peripheral here, given that internet shutdowns are not a primary threat in the U.S. context.

Given these priorities, platforms such as Bluesky and Mastodon are promising options. Both support federation, interoperability, and user- or instance-level moderation, satisfying the Critical and High-ranked requirements for self-governance and cross-instance communication. These platforms can be particularly effective when moderation and/or admin roles are held by participants of the worker community itself. For tech-savvy groups, custom implementations built on protocols such as the AT Protocol (with Personal Data Servers and custom clients), ActivityPub, or Nostr offer greater control and adaptability, but at the cost of increased setup complexity. 

This case study illustrates how the framework can help activists navigate the DSN space to identify platforms that support coordination, community building, and resilient self-governance under surveillance or political uncertainty.


\section{Discussion}
In an era of corporate tech capture, where centralized platforms disproportionately shape digital social life, users are seeking greater agency online. Repeated divestments from these platforms, often driven by policy shifts or ownership changes, reflect a growing public appetite for safer, more participatory alternatives---particularly among communities targeted by censorship, harassment, and surveillance. 
For activist communities, these alternatives must support secure coordination, exploration, sharing, and mobilization.

Decentralized social networks (DSNs) offer a promising path by giving users greater control and enabling communities to supplement or replace centralized platforms to meet their specific needs.
Yet, despite their potential, the DSN ecosystem remains complex and difficult to navigate. 
Our framework addresses this gap by breaking down the vast universe of DSNs, connecting activist community needs to concrete platform affordances.  
Rather than starting from protocol type or ideological stance, it offers a structured way for users to build mental models based on core needs: minimal overhead, community building and reach, on- and offline safety, and operational sustainability.  
This can ultimately guide communities in more intentional exploration and selection of platforms aligned with their priorities.
We also use the framework to explore, more broadly, how technical protocols and architectures influence---and are shaped by---community dynamics and governance. 

Our analysis shows that all DSNs prioritize certain values and affordances while constraining others, and that these tradeoffs have implications for how communities organize and sustain themselves. Community needs, too, are not uniform: while some groups prioritize visibility and growth through interoperable ecosystems and third-party integrations, others center safety and may thus deliberately limit content exposure. Navigating the DSN landscape, then, is a context-specific process that involves understanding which tradeoffs align with the goals and constraints of a particular collective. For example, we illustrate how internal communication and coordination may be best served by privacy-focused platforms, whereas outreach and engagement might benefit from networks with strong discoverability and integration capabilities. Our work builds on extensive CSCW literature examining how social media platform affordances support or constrain different community formations \cite{ghoshalRoleSocialComputing2019,harris2019joining,jiang2020characterizing,fieslerMovingLandsOnline2020}, extending this line of inquiry to the decentralized social ecosystem.

We also note that DSNs are not universal replacements for centralized platforms. Certain actions (e.g., mass mobilization) still rely on the scale and visibility that mainstream platforms offer. Nonetheless, the wide range of decentralized architectures and resulting community configurations highlight the critical role of infrastructure in shaping the potential for community autonomy.

\subsection{Limitations and Future Work}
\label{sec:limitations}

Our work provides a structured conceptual foundation for understanding the complex landscape of DSNs through the lens of the core needs of organizing communities. 
It aligns with other CSCW social media scholarship that presents conceptual frameworks that have been valuable foundations for the CSCW community, e.g., inheriting from affordance-based approaches~\cite{devitoPlatformsPeoplePerception2017}, employing validation methods beyond user studies~\cite{zhangFormFromDesignSpace2024a}, and centering the role of social media in community organizing~\cite{ghoshalRoleSocialComputing2019}. Like these precedents, our framework aims to advance research by providing a structured model for understanding how decentralized social networks afford activist organizing, enabling researchers and practitioners to systematically relate platform capabilities to grassroots community needs.

\subsubsection{An Evaluation Plan for the Framework}
A limitation of this work is that the framework has not yet been evaluated and refined in collaboration with activist communities.
We see empirical validation as an important next step.

We propose evaluating the framework against four criteria with both technical and domain experts. \textit{Completeness} asks whether the community needs and affordances in the framework capture activist priorities without critical gaps. \textit{Validity} asks whether the mappings between needs and affordances are accurate. \textit{Measurability} asks whether the features that operationalize each affordance are defined clearly and can be applied consistently across platforms. Finally, \textit{Utility} asks whether activists and organizers find the framework to be a useful lens for building a mental model of and reasoning about the DSN space.

To assess these criteria, we propose two validation phases. In the first, we plan to conduct semi-structured interviews with technologists specializing in DSNs and social computing, with a focus on completeness, validity, and measurability. In the second, we plan to engage activists directly through focus groups centered on their specific organizing contexts, to assess whether the framework captures their priorities and is legible across varied technical backgrounds. Because activists are the framework's intended users, this second phase is critical, particularly for assessing utility.

Insights from both phases would help us understand the framework's strengths and limitations, and highlight areas for iterative refinement. 
We also anticipate that evaluation with activists may surface context-specific nuances---different organizing situations require different platform capabilities---and we aim for the framework to preserve that interpretive flexibility rather than prescribe a specific ranking of platforms.

\subsubsection{Opportunities for Future Work}
We also see opportunities for future work with activist communities to support exploration and decision-making around DSNs in practice. While the framework provides a valuable conceptual foundation, translating it into specific platform choices still requires considerable effort. 
We created ``platform cards'' to illustrate the framework's ability to capture granular distinctions in technical affordances across different platforms, but we recognize that, in their current form, these cards function primarily as an analytic and summarization tool rather than a community-facing guide.
Future work could collaborate with activist communities to extend the framework and platform cards into practical resources---such as guides, visual artifacts, or interactive tools---that help groups understand key dimensions and make informed platform choices.
We point to similar conceptual contributions such as Model Cards \cite{mitchellModelCardsModel2019} and Datasheets \cite{gebru2021datasheets}, which, through distilling key sociotechnical dimensions of models and datasets, provided a foundation for subsequent work with stakeholders to develop visual artifacts \cite{pushkarnaDataCardsPurposeful2022}, interactive implementations \cite{crisan2022interactive}, application-specific adaptations \cite{rostamzadehHealthsheetDevelopmentTransparency2022}, and evaluations of their usage in practice \cite{nunes2024using}. Similarly, future work could build on the framework to co-design tools for navigating the DSN landscape---drawing inspiration from the searchability of \verb|instances.social|\footnote{https://instances.social/}, content presentation of \verb|DAO-Analyzer|~\cite{arroyoDAOAnalyzerExploringActivity2022}, or game-based design of \verb|Equality Engine|~\cite{showkatEqualityEngineFostering2025}.

Our analysis also raises conceptual questions about how digital infrastructures shape online community configurations. In Section \ref{sec:comm}, for example, we begin to draw out recurring patterns in tradeoffs across platforms. Are there certain combinations of affordances that are always mutually exclusive? Are there sets of user needs that are not concurrently satisfied by any available DSN? Can we further understand how and why distinct community configurations take shape across different DSN infrastructures? 
We hope that by formalizing key affordances, our work opens the door for future research to investigate exactly these kinds of questions. 

Finally, as we continue to engage with the complexity of the DSN landscape, we ask: who is truly able to participate in these spaces, and what structural barriers still limit more equitable adoption?  Future work should address the technical barriers to broader adoption of DSNs. This includes developing user interfaces that not only simplify interactions but also support learning about the underlying technologies. And while our focus has been on how activist communities might join existing decentralized platforms, there remains untapped potential to support their co-creation of features and systems---enabling them to actively shape the tools they rely on.

\subsection{Power, Inequality, and the Limits of Decentralization}

Decentralized social technologies are often heralded as vehicles for radical change---capable of reshaping political, economic, and social structures through their technical foundations~\cite{allen_ethics_2023, armano_emerging_2022}.
Advocates suggest these systems are inherently positioned to produce more equitable and participatory outcomes.
However, we have seen that the participatory affordances of current decentralized platforms vary widely, and that the interests, survival, and well-being of decentralized systems often do not align with those of their users ~\cite{frost-arnold_beyond_2024}.

To understand why, we must consider the broader sociotechnical conditions shaping DSN design and adoption. Despite some spaces---such as the fediverse, where protocols like ActivityPub were developed by queer and trans technologists---offering more inclusive foundations~\cite{hwangTrustFrictionNegotiating2025}, the broader DSN landscape remains dominated by developers and users from privileged demographics in the Global North. This results in a lack of diversity in terms of geography, culture, and gender~\cite{albusays_diversity_2021, subramanyam_freelibre_2008, newton_understanding_2024}.

Infrastructure disparities further exacerbate this imbalance. The physical and digital resources required to host and access DSNs---such as servers, high-performance devices, and stable power and internet access---are disproportionately concentrated in Western countries~\cite{denardis_3_2022}. The promise of decentralized networks as fully open and empowering alternatives is likely to remain unfulfilled as long as these deeper inequalities persist. The potential of decentralized social networks will only be realized if we move beyond technical solutions and actively confront systemic forces---capitalist, patriarchal, and imperialist---that shape access, participation, and control in digital life.


\section{Conclusion}
In this paper, we examine how decentralized social networks (DSNs) can support the work of activist communities. We begin by identifying key architectural topologies and protocols and analyzing how communities have historically used social media to build collective identity and engage in activism.
Next, we outline the specific needs of activist communities in digital spaces and introduce a framework that maps these needs to the affordances of DSNs. We apply this framework to several popular decentralized platforms, examining how their underlying infrastructures shape the ways communities form, organize, and interact.
We ground our analysis in two case studies to illustrate how activist groups engage with these platforms in practice. We conclude by discussing the limitations of our framework, proposing directions for future research, and reflecting on the broader challenges of using decentralized technologies to meaningfully shift systemic power imbalances.

\bibliographystyle{ACM-Reference-Format}
\bibliography{main}

\appendix

\newpage
\section{Framework Summary Table}
\label{appendix}


\begin{table*}[h]
\centering
\caption{Community Needs, Affordances, Definitions, and Features}
\footnotesize
\renewcommand{\arraystretch}{1.1}
\begin{tabular}{p{1.5cm} p{3.0cm} p{4.0cm} p{5.0cm}}
\hline
\textbf{Community Need} & \textbf{Affordance} & \textbf{Definition} & \textbf{Features} \\
\hline

Minimal overhead 
& Account creation accessibility
& The ease of becoming a registered, participating user. 
& Multi-modal access (web, mobile, SMS, on- and offline); Few setup steps; Zero-verification signup; Accessibility (i18n, screen readers); Migration tools \\

& Interface familiarity
& Users can navigate the interface using prior experience.
& Established visual and interaction conventions; Minimal jargon \\

& Smart defaults
& Users can engage immediately, without configuration, due to reasonable system-set defaults.
& Pre-populated feeds; Default privacy settings; Auto- or suggested instance selection \\

& Hardware accessibility
& The platform is usable on low-resource devices.
& Low storage/compute footprint; Cloud-hosted infrastructure; Related documentation \\

& Financial accessibility
& Users can participate meaningfully without cost.
& Free core actions (post, follow, comment); Affordable premium features \\

Community building and reach
& Discoverability
& Users can find relevant content, communities, or other users.
& Visibility mechanisms for presented content (options for content feeds); Access modes (public/login); Search; Content synchronization \\

& Interoperability
& The platform exchanges content, identity, or data with other systems.
& Open protocols; Cross-posting and bridge tools \\

& Customizability
& Users can modify content, interface, and platform behavior beyond defaults.
& Multiple content formats; Edit/repost/delete; Profile and UI customization \\

& Interactivity
& Users can form relationships and respond directly to content.
& Follows, DMs, groups; Likes, comments, shares \\

On- and offline safety
& Data privacy
& Users can control the visibility of their profiles, relationships, and actions.
& Pseudonymity; Content visibility controls (for posts, replies, comments); End-to-end message encryption; Verification indicators \\

& Data ownership and control
& Users can determine how their data is stored, secured, migrated, and deleted.
& Self-hosting; Data export/download; Migration; Deletion \\

& Moderation
& Users can restrict certain content and accounts.
& Reporting; Blocks, mutes, filters; Shared rule lists \\

Operational sustainability
& Governance structures
& Decision-making authority and access rights are assigned among users, admins, and developers.
& Instance/platform privileges; Decision-making mechanisms (for proposal creation and voting) \\

& Platform-level support
& Users can rely on consistent assistance and maintenance on the platform.
& Core development resources; Issue and feature channels \\

& Transparency
& Operators openly communicate about infrastructure, policies, and processes.
& Open-source code and APIs; Governance and policy documentation \\

\hline
\end{tabular}
\end{table*}

\end{document}